*Review Article*

# A Systematic Literature Review of Blockchain Technology Adoption in Bangladesh

**Abdullah Al Hussain, Md. Akhtaruzzaman Emon, Toufiq Ahmed Tanna, Rasel Iqbal Emon and Md. Mehedi Hassan Onik\***

American International University-Bangladesh (AIUB), Dhaka, Bangladesh
abdullahalhussainf11@gmail.com; azaman.emon@gmail.com; toufiqtanna@gmail.com; ras.iqbal.97@gmail.com; mehedi.onik@aiub.edu
\*Correspondence: mehedi.onik@aiub.edu

**Abstract:** The spirit of "blockchain technology" is a distributed database in which saved data is transparent, accountable, public, immutable, and traceable. This base-level disruptive technology can boost the security and privacy-related efficiency of various domains. As Bangladesh is currently aiming for sustainable development, blockchain technology adoption by the local researchers is growing robustly. However, in Bangladesh, the blockchain Technology Acceptance Model (TAM) is not yet well structured which is also limiting the perspective of local developers and researchers. Therefore, sectors like governance, healthcare, security, privacy, farming, information authentication, cryptocurrencies, internet architecture, data, and so on are unable to utilize the full potential of this technology. In this research, the authors conduct an in-depth review of such types of blockchain technology-related research articles that have been published recently and are also solely focused on Bangladesh. From 5 publishers (IEEE Xplore, ACM, ScienceDirect, Taylor & Francis, and SpringerLink) this study analyses 70 articles published during the year 2016-2020. The study results find the top 13 sectors where Bangladeshi researchers are currently focusing on. Those studies identify that the rigid policy by the government, scarcity of expert researchers, and lack of resources are the main reasons why Bangladesh is still struggling to accommodate blockchain extensively. In addition, published papers are mostly based on theoretical concepts without an appropriate implementation. Finally, this study will be a great resource to the developers, entrepreneurs, and technology enthusiasts to determine the strategic plan for adopting blockchain technology in Bangladesh or even to any other developing country.

## 1. Introduction

Satoshi Nakamoto initially introduced both Blockchain Technology and Bitcoin in 2008, where he defined how a cryptology-based decentralized and distributed public ledger can be accumulated into a digital application [1]-[2]. In the current policy, currency dealings between two persons or firms are centralized and measured by a third-party unit such as a bank [3]. In addition, to perform that transaction, a fee is also charged. To solve this aforesaid issue, blockchain technology has developed. But the concept of blockchain technology is not limited to digital exchange and finance anymore. It has regularly stretched into healthcare, management of supply chain, marketplace monitoring, intelligent energy, personal data privacy, and so on [4]–[7]. Different kinds of domains are going up with the importance of blockchain. As a developing country, blockchain will have a significant influence on it. If Bangladesh is allowed to use cryptocurrency, the corruption rate will be decreased dramatically [8]. In Bangladesh, there is limited progress in blockchain technology due to the lack of structures and directions. The current policy of the Bangladesh Bank, which deliberates bitcoin and additional cryptocurrencies to be illegitimate under the Foreign Exchange Regulation Act (FERA) of 1947 and the Money Laundering Prevention Act (MLPA) of 2012, is the most severe question[1]. In some other domains like Supply Chain, Healthcare, Business, Privacy, IoT, Data Management Areas, Proposed National Blockchain Strategies, Blockchain technology can also be executed [9]. In this document, there are some approaches that the authors tried to prescribe a pathway to become a blockchain-enabled nation. However, to achieve that, we need a team of Blockchain Experts, some Technologists who understand blockchain technology, academicians, government officials, and other collaborators with a better policy that supports blockchain-based technology. Our neighbour country India has progressed in many sectors in blockchain technology. Many countries have also accepted blockchain technology sectors earlier and now leading the entire world, such as China, Australia, Japan, UAE, Malta, Switzerland, USA, Estonia, U.K., Singapore[2].

In this study, we outline a systematic literature review on blockchain technology development, especially in Bangladesh. A systematic literature review defines, chooses, and objectively appraises a study to answer an articulated question [10]. It is essential to give a country the direction towards future researches, broadens the knowledge on the research topics, and also to detect which domains of research are inescapable that need to perform. Health systematic literature review [11], procedures for performing systematic reviews [12], software engineering [13], cloud computing service composition [14], microcredit industry [15], enhancing security through software-defined networking-IoT enabled architecture [16], are some good examples of systematic literature reviews that previously helped us while adopting with those sectors.

The main theme of this paper is based on the collection of some Statistical Data and Analysis Results which are mainly being focused on adopting blockchain technology within Bangladesh. We have collected data from five publishers like IEEE Xplore, ACM, Springer, Taylor & Francis, and ScienceDirect. We also have devised Blockchain-related journals and their distribution by discipline. Consequently, a total of 70 journal papers could be possible to find out by us that have prescribed mainly based on Bangladesh. To the best of our knowledge, this systematic literature review would be the first where a detailed analysis is done exclusively on the blockchain-related articles formed in Bangladesh.

### 1.1. Key Research Contributions

- To find out the progress of blockchain technology in Bangladesh.
- To ponder about the policy on blockchain technology in Bangladesh, policy-makers, educational institutions, and responsible technical persons.
- To unleash possible technical sectors of Bangladesh where we can use blockchain technology.
- Expressing all the barriers in Bangladeshi society during the adoption of blockchain technology.
- Searching those types of people who are working on executing blockchain technology in Bangladesh precisely.
- Finding out the papers and the domains that are being practiced on Blockchain Technology in Bangladesh.

## 2. Proposed Method

A detailed description of the Systematic Literature Review (SLR) of blockchain technology adoption in Bangladesh has illustrated in this section. An SLR is a formal procedure where researchers design and develop the research questions based on evidence [17]. This section provides insight on how blockchain technology is being evolved here in Bangladesh with adequate scientific evidence. It also delivers detailed information on the ongoing research domains, challenges, and a preamble place for future blockchain technology-related research scope in Bangladesh. Overall, the study critically examined blockchain technology-related research articles. The detailed SLR procedure has been adopted to Kitchenham's procedure that includes sections like research questions, article selections, manual selections, attributes framework, and articles assessment (Figure-1) [18]–[20]. Besides, this article also accommodates guidance to make systematic literature from another study as well [21].

---

[1] https://www.thedailystar.net/opinion/news/digital-bangladesh-needs-digital-currency-1930845 #:~:text=Bangladesh is one of the, Money Laundering Prevention Act%2C 2012 (accessed Mar. 30, 2021)
[2] https://www.blockchain-council.org/blockchain/top-10-countries-leading-blockchain-technology-in-the-world

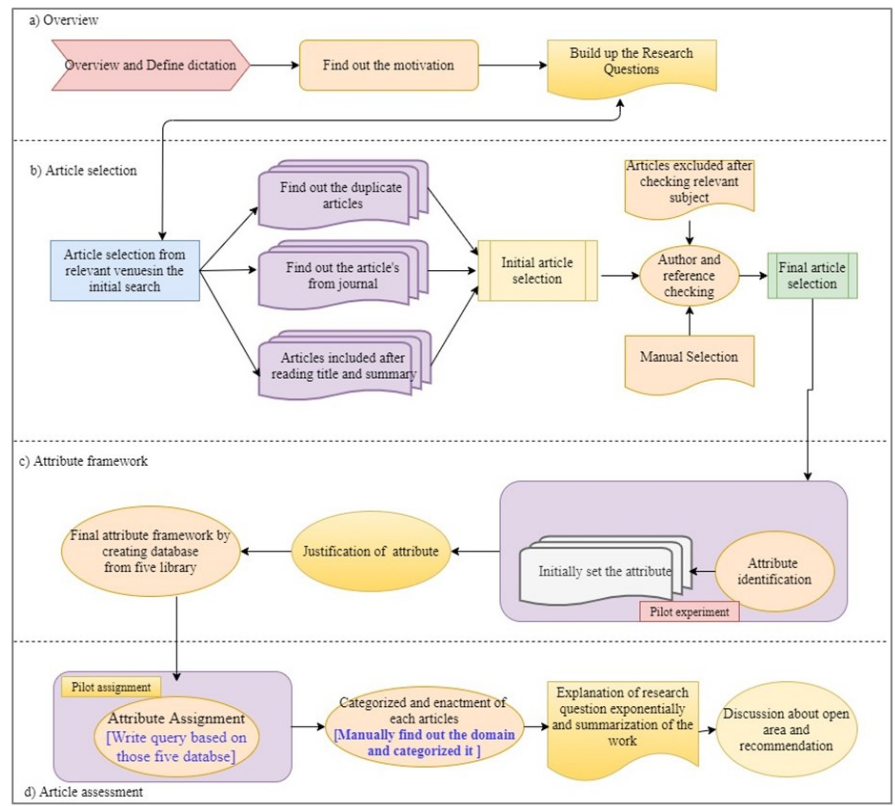

**Figure 1.** Overview of the Systematic Literature Review (SLR) Protocol.

Initially, this section provided a brief idea about the specified four sections demonstrated in Figure-1:

a) **Overview:** In the overview part, this study searches for the precised research directions and motives. Once the direction and motive selection is done, the study does decide to focus on the SLR of the blockchain technology adoption in Bangladesh. Finally, in this section, research questions based on the selected topics are done.

b) **Article Selection:** After a thorough analysis of the research questions, the study covers the initial searches for all the Blockchain Technology-related papers which address Bangladesh (adopted by the Bangladeshi authors or published in the Bangladeshi venues as well). Once the initial article selection has been done, a manual selection for inclusion and exclusion of articles has been completed to choose the final set of articles for further study.

c) **Attribute framework:** The attribute identification from the finally selected articles is evaluated by the pilot experiment. By performing this pilot experiment, this study has outlined the justified attributes. Finally, the framework tries to finalize the attribute and creates a database from the selective digital libraries for articles.

d) **Article assessment:** Article assessment is the final part of the SLR protocol. This part tries to finalize the pilot assignment by writing the query based on those selected databases. After writing all the queries, the study categorizes and examines each article manually to identify the domain. Then the research questions have been explained by this work exponentially and the work also has been summarized to reach our goal consecutively.

### 2.1. Goal and Research Questions

It has almost been a decade since Nakamoto introduced blockchain technology [22]. Afterward, countries all over the world are adopting this ground-breaking technology for multiple purposes. Depending on the applications, technological strengths, and research implications, every country has managed blockchain differently. Similarly, being a developing country, Bangladesh has also accommodated this blockchain technology with enormous opportunities for scientific and technological growth. Therefore, this study critically analyses Bangladesh-related blockchain technology articles to provide a concrete research direction and upcoming adoption challenges aiming especially all the policymakers, researchers, and government. For this, this study formulates three research questions (Table-1).

**Table 1.** Research Questions

|  | Research Questions |
|---|---|
| RQ1 | What are the application divisions in which blockchain adoptions are being utilized or developed in Bangladesh? |
| RQ2 | What are the impediments and encounters of blockchain technology adoption in Bangladesh? |
| RQ3 | How Policymakers should approach to adopt the blockchain technology for sustainable development in Bangladesh? |

### 2.2. Article Selection

In this article selection section, this study finalizes the articles taken into consideration for this systematic literature review (SLR). Five major millstones are shown in the review protocol (**Figure-1**) to complete this whole article selection procedure. In addition, this study also searches for some relevant publication venues followed by the identification of duplicate articles. Search results from several publishers were analysed based on their titles. Besides, a summary of the chosen articles was also created for future use through this study. Based on that initial article selection, this study focuses on the authors' nationality. Then, the articles are excluded or included after checking references before finalizing the ultimate article selection.

### 2.3. Inclusion Criteria (IC)

All the research questions and the domain of those researches that have been created based on the blockchain technology adoption in Bangladesh are mainly focused on this Inclusion Criteria. By following these inclusion criteria (IC), this study identifies the article selection condition that allows us to identify the research directions and challenges of blockchain use-case. (Table-2) illustrates the IC of this SLR.

**Table 2.** Inclusion Criteria (IC) of the Systematic Literature Review

| Serial Number | Description of Inclusion Criteria |
|---|---|
| IC1 | The selected papers must have strong relevance with blockchain technology which is adopted in Bangladesh, it should be verified from the title, abstract, and keywords from the article selection. |
| IC2 | Those articles that are already published in journals and conferences in Bangladesh should be selected for review. |
| IC3 | Those published articles written by Bangladeshi authors and those should be written in English. |
| IC4 | Should have collected the articles that have been published "between" 2016 to 2020. |
| IC5 | The selected papers must be blockchain technology-related but they can be accepted from various domains. |
| IC6 | Manual analysis is imperative for picking up the conceivable significant papers accurately concerning this subject with the assistance of title, keywords, and abstract. |

### 2.4. Publishers' Library Selection

This study uses five libraries for searching the relevant research topics. From these publishers' libraries, this work selects some chapters of books, journal articles, conference articles that are relevant to the study topics which are involved in a systematic literature review of blockchain technology adoption in Bangladesh. By using the search options of each publication library, this study critically scrutinizes to include the articles related to our study objective. To be more specific, the searching factors are confined to Title, Abstract, and keywords. All the five publishers' libraries are listed in (Table-3).

**Table 3.** List of Libraries Searched

| No. | Library |
|---|---|
| 1. | IEEE Xplore |
| 2. | Association for Computing Machinery (ACM) |
| 3. | ScienceDirect |
| 4. | Taylor & Francis |
| 5. | SpringerLink |

**2.5. Query Processing on Database**

We make a query processing on those five databases (Table-4). We started queries on the first of March in 2021. By doing this query we found 70 articles researched by the Bangladeshi authors (Figure-2).

Table 4. Querying processes within publishers' databases

| Search criteria | IEEE Xplore | ACM | ScienceDirect | Taylor & Francis | SpringerLink |
|---|---|---|---|---|---|
| Query Date | 17 March 2021 | 17 March 2021 | 17 March 2021 | 17 March 2021 | 17 March 2021 |
| Query | (((("All Metadata": blockchain) AND "All Metadata": technology) AND "All Metadata": Bangladesh) AND "All Metadata": systematic literature review) (("Abstract": blockchain) AND ("Abstract": technology) AND ("Abstract": Bangladesh)) | (((Abstract: blockchain) AND (Abstract: technology) AND (Abstract: Bangladesh) AND (Abstract: systematic literature review)) "filter": ACM Content: DL) (((Abstract: blockchain) AND (Abstract: technology) AND (Abstract: Bangladesh)) "filter": (ACM) Content: (DL)) | (((("All Metadata": blockchain) AND "All Metadata": technology) AND "All Metadata": Bangladesh) AND "All Metadata": systematic literature review) | ((Abstract: blockchain) AND (Abstract: technology) AND (Abstract: Bangladesh) AND (Abstract: systematic literature review)) | (((("All Metadata": blockchain) AND "All Metadata": technology) AND "All Metadata": Bangladesh) AND "All Metadata": systematic literature review) FROM (Computer Science (Information Systems Applications) AND (Systems and Data Security)) |
| Papers found | $Q_{ieee} = 37$ | $Q_{acm} = 8$ | $Q_{sciencedirect} = 3$ | $Q_{taylor\ \&\ francis} = 1$ | $Q_{springer} = 21$ |

**2.6. Domain Evaluation**

In this study, we have separated the specified domains into three parts including Technical issues and solutions related articles, Policy and Recommendation related articles, and eventually Overview and other articles. In the entire paper, we have found 70 articles and we separate them into these 3 main domains (Technical issues and solutions related Articles, Policy and recommendation related Articles, Overview and other relevant Articles) (Figure 3). Afterward, multiple sub-domains are formed based on the aforesaid 3 main domains (Table-5) (Figure 5). The study then considers only the top 13 sub-domains (Table-5) (Figure 5).

Table 5. Description of selected top 13 sub-domains of available articles

| Sub-Domain | Details Description of Sub-Domain Selection Criteria |
|---|---|
| Governance | The articles are selected as governance domain because these articles are related to government works or maintain the rules of the government. Besides that, those articles are related to different organizational security purposes are selected as security domains. |
| Privacy | The articles are selected as privacy domains because these articles are identity or privacy-related papers of blockchain technology. Besides that, those articles are related to organizational privacy and security purposes. |
| Storage | The articles are selected as storage domains because these articles are related to storage papers of blockchain technology. |
| Architecture | The articles are selected as architecture because these articles are related to the architectural programmed related paper in blockchain technology. |
| Healthcare | Healthcare domain-related papers are describing the practice of blockchain technology in the healthcare sector in Bangladesh. |
| Survey | The articles are selected as survey and review domains that are the literature review or survey time paper which all are published in Bangladesh. |
| Farming | The articles are selected as farming domains that are related to farming using blockchain technology. |
| Security | The articles are selected as security domains because these articles are identity or secrecy-related papers of blockchain technology. |
| Review | The articles are selected as review domain, is the literature review paper which all are published in Bangladesh. |
| Information Verification | The articles are selected as information verification domain, this verify the information and the paper which all are published in Bangladesh. |
| Cryptocurrencies | The articles are selected as Cryptocurrencies domain, these articles are worked on Cryptocurrencies technology and the paper which all are published in Bangladesh. |
| Internet | The articles are selected as Internet domain, these articles are worked on internet and the paper which all are published in Bangladesh. |
| Data | The articles are selected as data, these articles are worked on data and information and the paper which all are published in Bangladesh. |

## 3. Results and analysis

### 3.1. Overview

As a result of this survey, we found a total of 70 papers that are published in different five publisher websites as represented in the following graph. In the result, **IEEE Xplore** is the biggest contributor with 37 papers. On contrary, from **Taylor & Francis**, we just could collect only 1 paper (Figure-2).

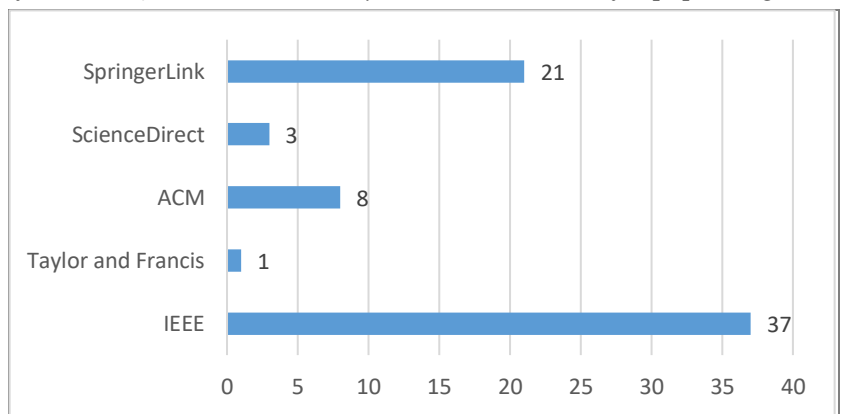

**Figure 2.** Graph for the overview of the Systematic Literature Review (SLR) Protocol.

In Table (6-10), we have tried to find out the domain specifications for each paper we collected. We have read each paper's abstract and title and then from that knowledge we have divided the analysis into 3 major domains (Figure-3). Moreover, we divided them into 13 sub-domains (Figure-4) and eventually tried to provide a brief description on the key concepts of each article.

**Table 6.** Overview of Articles considered in this Systematic Literature Review (IEEE Xplore)

| Articles | Major Domain | Sub-Domain | Key Concept |
|---|---|---|---|
| Vehicle Registration and Information Management using Blockchain based Distributed Ledger from Bangladesh Perspective [23]. | Technical Issue and Solution | Governance | The system is capable of information management related to vehicle registration without a direct third parity involvement, unlike the paper-based system. |
| A Novel Framework for Implementation of Land Registration and Ownership Management via Blockchain in Bangladesh [24]. | Technical Issue and Solution | Governance | Proposes an improved method to overwhelmed manual land registration processes problems using blockchain technology and giving true and unquestionable rights on proprietorship for the individuals in Bangladesh. |
| Secure and Transparent KYC for Banking System Using IPFS and Blockchain Technology[25] | Policy and recommendation | Security | The proposed framework permits a client to open an account at any of a Bank, total the KYC handle there, after creating hash rate, utilizing the IPFS arrange and portion it using the blockchain method. |
| A Novel Framework for Blockchain Based Driving License Management and Driver's Reputation System for Bangladesh [26] | Technical Issue and Solution | Governance | Key highlights of blockchain and smart contracts created an improved driving license administration, driver's reputation framework, and computerized |
| | | | a few repetitive steps for a proficient system. |
| SDoT-NFV: A Distributed SDN Based Security System with IoT for Smart City Environments [27] | Technical Issue and Solution | Privacy | A secure SDOT-NFV design meant smart cities have been proposed. The architecture safeguards metadata within each layer and the payload. |
| DistB-CVS: A Distributed Secure Blockchain based Online Certificate Verification System from Bangladesh Perspective [28] | Technical Issue and Solution | Governance | The proposed "DistB-CVS" aims to close problems in the prevailing systems to authenticate academic record validity. |
| Blockchain based Land Registry with Delegated Proof of Stake (DPoS) Consensus in Bangladesh [29] | Technical Issue and Solution | Governance | An approach to upgrading the assigned confirmation of stick agreement to supply a unique ledger-based structure for executing land properties that can be effectively coordinated into the current conventional Land Registry system for easier procedure. |
| Telemedicine System Design using Blockchain in Bangladesh [30] | Technical Issue and Solution | Governance | This study proposed a demonstration of a telemedicine system utilizing blockchain innovation for the medical care of provincial people. It'll offer assistance to those people with proper medical care with the security of private information. |
| Increasing Throughput and Reducing Storage Bloating Problem Using IPFS and Dual-Blockchain Method [31] | Technical Issue and Solution | Storage | Paper is persuaded due to the need for versatility on the prevailing blockchain technologies. A conveyed capacity framework IPFS is utilized to avoid putting absent obligations and to expand output. |
| Efficient Blockchain System based on Proof of Segmented Work [32] | Overview and others | Architecture | This system is attempted to decrease energy utilization by restricting the interest of all the hubs within the network by putting a parcel of emphasis on the concept of fair reward distribution which isn't considered in most cryptocurrencies. |
| Towards Blockchain-Based E-voting System [33] | Technical Issue and Solution | Governance | An internet of things (IoT) constructed framework is planned to exchange information from e-voting gadgets to the nodes and the approach uses a conveyed ledger technology where information is shared and conveyed into a network. |
| Block-SDoTCloud: Enhancing Security of Cloud Storage through Blockchain-based SDN in IoT Network [34] | Policy and recommendation | Security | Block-SDoTCloud planning to essentially improve sanctuary inside the cloud capacity arrange by centralizing SDN controller given the proposed show with huge unwavering quality, flexibility, and load adjusting. |
| Smart Grid Implementation with Consortium Blockchain: A Proposed Model for Bangladesh [35] | Technical Issue and Solution | Security | The proposed smart grid will be able to guarantee operator safety with the assistance of advanced innovations and illuminate information susceptibility subjects and protect against refusal of facilities. |
| DistB-Condo: Distributed Blockchain-Based IoT-SDN Model for Smart Condominium [36] | Overview and others | Security | The approach is to beats the prevailing OF-Built SDN. DistB-Condo has a way of recovering output on normal, and the transfer speed (Mbps) is much advanced |

| | | | than the OF-Based SDN method within the nearness of assaults. Moreover, the projected demonstration has a normal reaction period fewer than the center demonstration. |
|---|---|---|---|
| A Hyper-ledger Fabric Framework as a Service for Improved Quality E-voting System [37] | Technical Issue and Solution | Governance | The foremost protected way of chosen by-election with the assistance of Hyper-ledger Fabric-built System as a Benefit (FaaS) which can be secondhand to execute an exceedingly sustainable. Its arrangement with a private blockchain. |
| Permission-Based Blockchain with Proof of Authority for Secured Healthcare Data Sharing [38] | Policy and recommendation | Healthcare | Blockchain with PoA innovation will ensure information security, information owner's control on sharing delicate data. It also addresses the significance of overseeing medical records in an interoperable way FHIR helps in the important exchange of healthcare records among all members. |
| A feedback system using blockchain technology [39] | Policy and recommendation | Survey | A digital feedback system will be utilized where input will be guaranteed with the namelessness of the users and also make sure the belief of users is being stalled when feedback is given. Moreover, it'll grant a thought of how the people of an organization think about certain themes. |
| A Critical Review of Concepts, Benefits, and Pitfalls of Blockchain Technology Using Concept Map [40] | Policy and recommendation | Survey | A systematic literature survey method was embraced in this study in the direction to accomplish the impartial. Total 51 articles were chosen and surveyed. As an outcome, this investigation gives an outline of the advanced inquires revisions conducted within the part of the blockchain. |
| An Approach For an Distributed Anti-Malware System Based on Blockchain Technology [41] | Technical Issue and Solution | Security | A distributed system of Antimalware database management utilizing customized blockchain that improves system security by starting conveyed malware avoidance of program. blockchain can guarantee better information management without including any third party. |
| SmartBlock-SDN: An Optimized Blockchain-SDN Framework for Resource Management in IoT [42] | Technical Issue and Solution | Security | A disseminated effective blockchain-allowed SDN-IoT system that confirms productive cluster-head assortment and sheltered organize announcement. The proposed framework can accomplish improved vitality utilization, endways adjournment, and quantity associated with measured baselines being able to secure within the shrewd network. |
| A Secured Smart National Identity Card Management Design using Blockchain [43] | Technical Issue and Solution | Privacy | Proposed a prototype for a smart NID card management framework utilizing blockchain innovation. blockchain earned ubiquity among analysts to ensure information from illegal quarrels so adjusting a prototype will support the |

| | | | administration of Bangladesh to protected citizens' isolated data and convey transparency to people's data administration. |
|---|---|---|---|
| Block - NID: A Conceptual Secure Blockchain Based National Identity Management System Model [44] | Technical Issue and Solution | Privacy | Attending to analysis mutually the current and most recent improvements in the arena of ID card framework by applying blockchain as a prototypical where moreover talk about the applications of blockchain besides the challenges confronted and future points of view. |
| FPoW: An ASIC-resistant Proof-of-Work for Blockchain Applications [45] | Overview and others | Architecture | A Filtered Proof-of-Work (FPoW) is projected and ASIC assessed to create an impending-impervious ASIC-resistant agreement convention. |
| A Secure IoT Data Communication and Sharing Framework Based on Permissioned Blockchain [46] | Technical Issue and Solution | Security | A blockchain-based system for IoT gadget information sharing and communication was actualizing a model of our proposed system utilizing Hyper ledger Fabric and assessed its execution and the explanation is effective for IoT gadgets and diverse safekeeping necessities for example privacy, accessibility, and judgment of information are encountered. |
| Digital Voting: A Blockchain-based E-Voting System using Biohash and Smart Contract [47] | Technical Issue and Solution | Governance | the study combines digitalization with blockchain innovation to supply a voting apparatus. Most areas of the elective machines are to supply astuteness, secrecy, protection, and haven of electorates. |
| Integrating Blockchain with Artificial Intelligence for Privacy-Preserving in Recommender Systems [48] | Technical Issue and Solution | Privacy | Private-Rec, a privacy-preserving stage aimed at a proposed framework finished the combination of AI and blockchain. In Private-Rec, blockchain stretches the shopper a safe situation through the dispersed characteristic in which information can easily utilize with the specified consent. |
| Modelling Attacks in Blockchain Systems using Petri Nets [49] | Technical Issue and Solution | Security | Investigate dissimilar exposures in present blockchain classifications and analyses the dangers that different hypothetical and commonsense assaults within the blockchain depiction. At that point demonstrate those assaults utilizing Petri nets regarding present frameworks and upcoming significant processers. |
| Towards SDN and Blockchain based IoT Countermeasures: A Survey [50] | Policy and recommendation | Review | This paper is proposed the observe IoT system features, sanctuary conditions, and encounters. From that point, renowned dangers of assaults are studied in IoT. SDN and blockchain-based remeasure are talked for IoT to organize sanctuary with a circumstance. |
| Secure Data Transaction and Data Analysis of IoT Devices Using Blockchain [51] | Technical Issue and Solution | Security | Studies how the blockchain arranges is made and period idleness of distribution |
| | | | the information of the IoT gadgets concluded this blockchain organize and a few parts of the examination handle which is essential for airship support. |
| A framework for city wide activity data recorder and providing secured way to forensic users for incidence response [52] | Policy and recommendation | Privacy | A system has been planned, which can preserve a blockchain-based record framework to avoid information interfering, a blockchain-built information assembly has been castoff by the system to guarantee permanence. |
| Distblockbuilding: A Distributed Blockchain-Based SDN-IoT Network for Smart Building Management [53] | Technical Issue and Solution | Security | Design a sensible building framework, together with a sway framework is projected. a good cluster head choice rule is projected to decide on the required cluster head. "DistBlockBuilding" design is dead for overseeing a secure and protected information handover from one superficial to a different superficial. |
| Fake News Detection in social media using Blockchain [54] | Overview and others | Information verification | Fake news is more appealing than a genuine one. In this way, individuals got to be confused. Utilizing the preferences of blockchain's peer-to-peer arrange concepts identify fake news in social media. |
| An Unorthodox Way of Farming Without Intermediaries Through Blockchain [55] | Policy and recommendation | Farming | Projected a regionalized farming stage, called KHET to resolution the specified problem and establishes belief and regionalization amongst agricultural partners for example agronomists, source businesses, and marketplaces. |
| Electronic Medical Record Data Sharing Through Authentication and Integrity Management [56] | Technical Issue and Solution | Data | Electronic Restorative records or EMR are exceptionally delicate information for the patients and the patients and government organizations. These frameworks need huge interoperability and get to restrictions. Suggest an EMR framework that can regulator the supervision and capacity through blockchain |
| Prediction of Dengue Infected Areas using A Novel Blockchain based Crowdsourcing Framework [57] | Technical Issue and Solution | Healthcare | The Dengue Tracker Framework is based on a disseminated crowdsourcing system. Diseased affected role and the aware inhabitant can yield conceivable irresistible areas. the framework is conveyed on Ethereum-blockchain to improve the sanctuary of the framework and avoid false area information. |
| From Conventional Voting to Blockchain Voting: Categorization of Different Voting Mechanisms [58] | Technical Issue and Solution | Governance | This study focused on a systematic investigation of numerous shapes chosen by-election with blockchain and lacking blockchain by abundant analysts as a medium to discover liabilities or shortcomings and suggest an ideal chosen by election instrument to achieve all the assets necessitates in the voting framework. |
| A Hybrid Blockchain-based Zero Reconciliation Approach for an Effective Mobile Wallet [59] | Technical Issue and Solution | Security | Proposes a half-breed blockchain-based folder plan to allow a speedier conversation among endways hubs and resistor exchange disappointment. To create the blockchain adjustable for the case, got to empower quicker interactions and guarantee safety; in this way, somewhat imitation never is happened. |

**Table-7.** Overview of Articles considered in this Systematic Literature Review (ACM)

| Articles | Major Domain | Sub-Domain | Key Concept |
|---|---|---|---|
| Barriers to Growth of Renewable Energy Technology in Bangladesh: Case of Solar Home System in Rural Regions [60] | Overview and others | Governance | This investigation examines the boundary to RET development for creating a provincial charge and considers the solar home system (SHS) an instance. The consider comprised non-parametric trial and literature study. Two spread found towns since the areas of Noakhali and Sirajganj, chosen for any situation to consider. |
| Blockchain based Fertilizer Distribution System: Bangladesh Perspective[61] | Technical Issue and Solution | Governance | proposed a system with the utilizes of blockchain technology-based fertilizer dispersion supply chain administration process. The tamper-proof information capacity and client administration framework ensure diligently observing and assessment preparation with the solid administration data framework which utilizes exact reference data that can be related to the board. |
| Achievements and expectations of digital Bangladesh: e-governance initiatives in Bangladesh [62] | Policy and recommendation | Governance | Bangladesh has announced the "Vision 2021" to set up an ingenious and modern nation by 2021. The government has actualized a few activities in e-commerce website, e-business, e-finance in addition to the improvement of phone, or handset organize volumes. This study portrays the existing accomplishments and desires of Advanced Bangladesh concluded e-governance creativities. |
| E-governance initiatives in Bangladesh [63] | Policy and recommendation | Governance | Adjustment of e-Governance may be important for canny governance and producing information technology (IT) significant to conventional inhabitants in Bangladesh anywhere a huge extent of inhabitants endures from the numerical gulf. This study offerings a standard appraisal of prevailing measurements wants, and choices for actualizing e-Ascendency in Bangladesh. |
| A Cryptanalysis of Trustworthy Electronicvoting using Adjusted Blockchain Technology [64] | Technical Issue and Solution | Cryptocurrencies | Electronic voting is a productive strategy that seems to offer assistance to individuals who seem not to vote for a restricted period. It too solves the issue of long voting time due to numerous voters or more complicated voting steps so proposing cryptanalysis on a dependable electronic voting plot with |

| | | | blockchain and discovers out that there are some issues in the main proposed plot. |
|---|---|---|---|
| Understanding the software development practices of blockchain projects: a survey[65] | Policy and recommendation | Survey | This study points to investigate the program-building practices of these ventures. Computer program building strategies counting taxing and safekeeping superlative hones got to be adjusted more reality to discourse the single features of blockchain. |
| Secured Blockchain Based Decentralised Internet: A Proposed New Internet[66] | Policy and recommendation | Internet | Attempt to depict with blockchain technology the decentralization of the web. A decentralized arrangement that empowers the web to function from the smartphone or tablet of anyone rather than centralized servers. |
| A decentralized computational system built on Blockchain for educational institutions [67] | Technical Issue and Solution | Governance | This research aims to present a decentralized demonstration of the additional scheme constructed on blockchain for instructive teach, which can relieve a few downsides of a centralized system and r explore the conceivable outcomes of presenting a cryptocurrency inside the established organization. |

**Table-8.** Overview of Articles considered in this Systematic Literature Review (ScienceDirect)

| Articles | Major Domain | Sub-Domain | Key Concept |
|---|---|---|---|
| A Blockchain-based Land Title Management System for Bangladesh [68] | Technical Issue and Solution | Governance | A blockchain-based arrangement that provides information synchronization and transparency. Considering the technological information and capacity of the individuals and the government and presented a phase-by-phase blockchain selection show that begins with an open blockchain ledger and later continuously joins two levels of Crossbreed blockchain. |
| IoT-Cognizant cloud-assisted energy efficient embedded system for indoor intelligent lighting, air quality monitoring, and ventilation [69] | Technical Issue and Solution | Security | This study has endorsed an entrenched framework for self-directed illumination and airing framework. The framework is competent in classification factual-time information in the cloud. Announcement with the regulatory framework takings place utilizing the GSM and Wi-Fi arrange to utilize the present-day HTTP convention. |
| Privacy-friendly platform for healthcare data in cloud based on blockchain environment [70] | Overview and others | Privacy | A healthcare information administration framework utilizing blockchain technology as capacity assistance to accomplish security. Cryptographic purposes are utilized to scramble affected role information and to guarantee pseudonymity and analyzing the information handling methods additionally the fetched adequacy of the shrewd contracts utilized in our framework. |

**Table-9.** Overview of Articles considered in this Systematic Literature Review (Taylor & Francis)

| Articles | Major Domain | Sub-Domain | Key Concept |
|---|---|---|---|
| Improving cloud data security through hybrid verification technique based on biometrics and encryption system [71] | Technical Issue and Solution | Security | The essential objective of this study is to turn away data admittance from cloud data storing by unaccepted by the patrons. This exertion employs fingerprint and check the biometric scheme and progressed encoding ordinary as a reliable encoding framework. |

**Table-10.** Overview of Articles considered in this Systematic Literature Review (SpringerLink)

| Articles | Major Domain | Sub-Domain | Key Concept |
|---|---|---|---|
| Automated Tax Return Verification with Blockchain Technology [72] | Technical issues and solutions | Governance | An automated tax confirmation framework utilizing blockchain technology which can decrease the potential debasement in numerous folds. Update or deletion of any information in the blockchain framework isn't permitted. The inherent security include of the blockchain makes the information secure, tamper-proof, and troublesome to hack. |
| Cyber Threat Mitigation of Impending ADS-B Based Air Traffic Management System Using Blockchain Technology [73] | Technical Issue and Solution | Governance | This study presents the cybersecurity risk to flying operations that has ended up more exciting with the progression on ATM which may generally hinge on ADS-B. It influences the cooperative energy of mediator systems and affirmed blockchain innovation like Hyperledger Fabric and ensures the sheltered, strong, and combined organization for the association. |
| Towards developing a secure medical image sharing system based on zero trust principles and blockchain technology [74] | Technical Issue and Solution | Healthcare | A framework to contest medicinal information exposures has been projected. The framework brands utilize and try to unchanging nature of blockchain, the extra sanctuary of null belief standards, and the adaptability of off-chain information capacity employing Interplanetary File System (IPFS). |
| A Systematic Review for Enabling of Develop a Blockchain Technology in Healthcare Application: Taxonomy, Substantially Analysis, Motivations, Challenges, Recommendations and Future Direction [75] | Policy and recommendation | Review | This considers methodically looks all-important study on blockchain in health care app in different three available records. They think about utilized the characterized catchphrases 'blockchain', 'electronic health records' and 'healthcare' and also use few other things in their studies. |
| Blockchain-Based Information Security of Electronic Medical Records (EMR) in a Healthcare Communication System [76] | Policy and recommendation | Healthcare | This study presents a blockchain-built show for safeguarding Electronic Medical Records (EMR). Utilized SHA256 protected hash calculation for creating special indistinguishable 256-bit or 32-byte hash values for a specific therapeutic result. |
| Examining Usability Issues in Blockchain-Based Cryptocurrency Wallets [77] | Policy and recommendation | Security | Investigate common convenience topics with computer and cell phone-based |
| | | | cases and outcomes uncover that mutually cases need great convenience in acting the basic assignments which can easily be upgraded essentially and outline the discoveries and point out the viewpoints where the issues exist. |
| Towards a Blockchain-Based Supply Chain Management for E-Agro Business System [78] | Technical Issue and Solution | Governance | Suggest an effective, effective, and acceptable framework and benefit arrangement to agro-business dealers conjointly a nutrition traceability framework built on blockchain and by the help of IoT to form commercial smarter and also wealthier. Besides the blockchain and also the smart contract assistance of IoT beams, attempted to work on the most extreme exertion to diminish humanoid mediation. |
| IoT Security Issues and Possible Solution Using Blockchain Technology [79] | Technical Issue and Solution | Security | Suggests a novel strategy to protected IoT gadgets utilizing blockchain innovation. With the heterogeneous network, littler estimate, memory, capacity, and insecure estimation, this is simple to drudge and control the IoT gadgets. To have a protected and protected association, this could be much-executed things to guarantee information safety in IoT as well. |
| Integration of Blockchain and Remote Database Access Protocol-Based Database [80] | Technical Issue and Solution | Data | This study suggests, illustrates the utilize of blockchain for safeguarding little endeavors in contrast to hacking by alarming the administration at whatever point a alter is completed to the information deprived of utilizing the official networks. It can be completed through blockchain technology's inborn hash duplication and mining calculation. |
| A Blockchain-Based Scheme for Sybil Attack Detection in Underwater Wireless Sensor Networks [81] | Technical Issue and Solution | Security | A blockchain-built Sybil assault discovery plot in UWSN. They try to coordinate one of their past beliefs a show with the blockchain-based strategy to create it strong against the assault's discovery and conducted a test in Carafe and talked about the usage with code subtle elements. |
| Understanding the motivations, challenges and needs of Blockchain software developers: a survey [82] | Policy and recommendation | Survey | The essential impartial of this paper is to realize the inspirations, contests, and wants of BCS designers and study the contrasts between BCS and non-BCS improvement. The comes about to recommend that a big part of the BCS designers are skilled in non-BCS advancement and they are worked in spurred by the belief system of making a regionalized money-related framework. |
| A Combined Framework of Interplanetary File System and Blockchain to Securely Manage Electronic Medical Records [83] | Overview and others | Security | Projected a blockchain mutual with the Interplanetary File System arrangement system for EMR in health care |

| | | | manufacturing. The main point is to execute blockchain EMR and give get-to rubrics for different clients. The planned system, whereas securing quiet protection, permits helpful get to by endorsed specialists such as healthcare suppliers to therapeutic information. |
|---|---|---|---|
| A Novel Approach to Blockchain-Based Digital Identity System [84] | Technical Issue and Solution | Privacy | Projected a framework on blockchain-built advanced personality for people utilizing bio-information and executed proposed digital identity framework utilizing Ethereum savvy contract. |
| MediBchain: A Blockchain Based Privacy Preserving Platform for Healthcare Data [85] | Technical Issue and Solution | Healthcare | A healthcare information administration framework by utilizing blockchain as the capacity to accomplish secrecy. Artificially it is guaranteed and also utilizing the cryptographical capacities to secure affected role's information. |
| CRAB: Blockchain Based Criminal Record Management System [86] | Technical Issue and Solution | Governance | Presents a criminal record storage framework by actualizing blockchain technology to store the information, which makes a difference to accomplish astuteness and security this framework presents ways in which the authority can keep up the records of criminals effectively. |
| Blockchain-based security management of IoT infrastructure with Ethereum transactions [87] | Technical Issue and Solution | Security | This study carries frontward a modern authoritative, insinuate, and insubstantial brickwork for IoT-built blockchain innovation which leaves the memory above and central framework, whereas safety and sanctuary aids are preserving. |
| An E-Voting Framework with Enterprise Blockchain [88] | Technical Issue and Solution | Governance | Planned a plot of the E-voting framework grounded on authorization or undertaking blockchain. The arrangement planned here is reasonable for connected electives of minor to average-scale enterprises, understudy characteristic decisions in colleges, and leading a decision for proficient figures. |
| A Secured Electronic Voting System Using Blockchain [89] | Technical Issue and Solution | Governance | Projected a blockchain-built voting instrument which is based on the foreordained go for individually hub to excavations new pieces in the blockchain instead of execution excessive calculation to pick up the opportunity to mine a block and investigated two conceivable clashing circumstances and projected a settling instrument |
| Reducing Storage Requirement in Blockchain Networks Using Overlapping Data Distribution [90] | Technical Issue and Solution | Data | Propose a strategy to separate the entire blockchain of dealings into a few non-lapping ruins and a variety of numerous duplicates of their working process. At that point, distribute these shards consistently over the nodes within the organization. |

| Blockchain-Based Secure E-Voting with the Assistance of Smart Contract [91] | Technical Issue and Solution | Governance | In this study, a completely dispersed e-voting framework founded on blockchain technology is projected. This convention exploits smart contracts within the e-voting framework to contract with sanctuary issues, exactness, and electorates' protection throughout the voting time. |
| --- | --- | --- | --- |
| Secured Smart Healthcare System: Blockchain and Bayesian Inference Based Approach [92] | Technical Issue and Solution | Healthcare | This research, about giving a multi-layer sanctuary system for mixed IoHT frameworks utilizing Bayesian inference-constructed belief organization with the carefully contracted blockchain arrangement which keeps the incorporates person information assurance. |

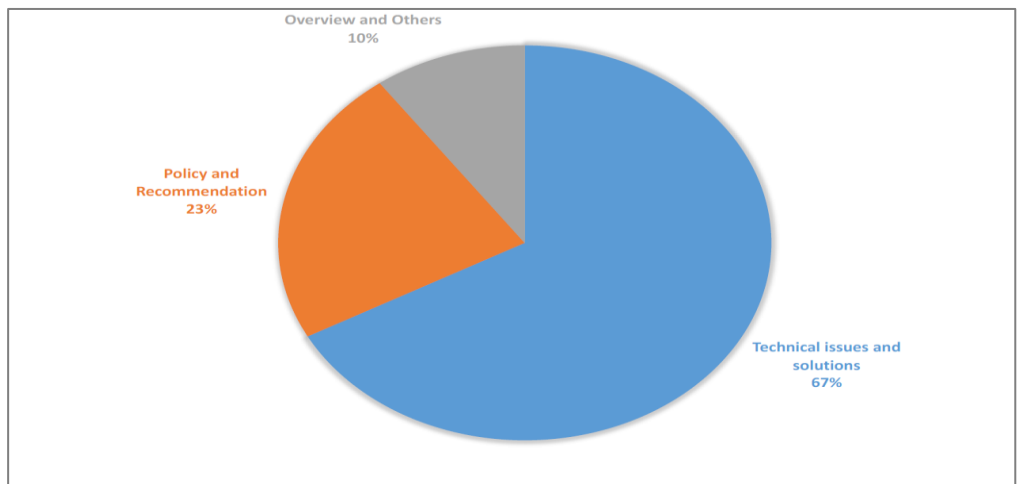

**Figure 3.** Major Domain.

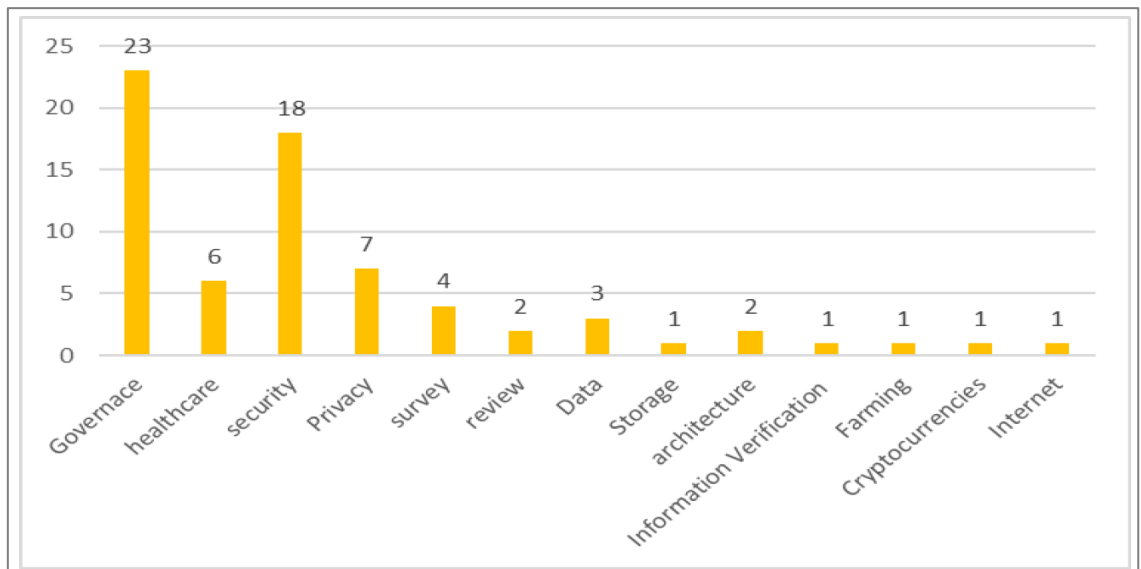

**Figure 4.** Sub-Domain.

### 3.2. Domain-Specific Statistics:

#### 3.2.1. Major Domain vs Publishers Library:

We got 47, 16 and 7 papers that are relevant to technical issues and also related to the solutions for technical issues, policy and recommendations, and Overview and others respectively (Figure-5). Regarding technical issues and solutions, we got 25 papers from IEEE Xplore, 1 from Taylor & Francis, 3 from ACM, 2 and 16 from ScienceDirect and SpringerLink respectively (Figure-5). For policy and recommendations, IEEE Xplore has 8 papers, Taylor & Francis has 0, ACM has 4, ScienceDirect has 0, and SpringerLink has 4 papers (Figure-5). From Overview and others, there are 8 papers in IEEE Xplore, 0 papers in Taylor & Francis, 4 papers in ACM, 0 papers in ScienceDirect, and 4 papers in SpringerLink (Figure-5).

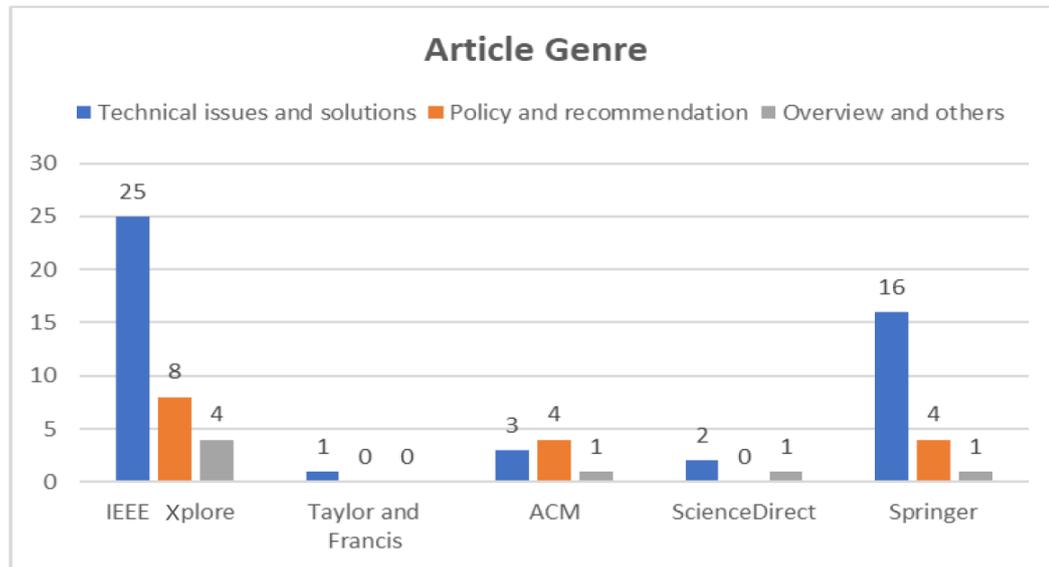

**Figure 5.** Article Genre

### 3.2.2. Sub-Domain vs Publishers Library:

In (Figure-4), we try to figure out the total 13 sub-domains papers number and the five databases holding papers numbers (Table-11).

Table 11. Distribution of selected top 13 sub-domain of available articles

| Sub-Domain | IEEE Xplore | SpringerLink | ScienceDirect | ACM | Taylor & Francis |
|---|---|---|---|---|---|
| Governance (23) | 10 | 5 | 1 | 7 | 0 |
| Healthcare (6) | 2 | 4 | 0 | 0 | 0 |
| Security (18) | 11 | 5 | 1 | 0 | 1 |
| Privacy (7) | 5 | 1 | 1 | 0 | 0 |
| Survey (4) | 2 | 1 | 0 | 1 | 0 |
| Review (2) | 1 | 1 | 0 | 0 | 0 |
| Data (3) | 1 | 2 | 0 | 0 | 0 |
| Storage (1) | 1 | 0 | 0 | 0 | 0 |
| Architecture (2) | 2 | 0 | 0 | 0 | 0 |
| Information Verification (1) | 1 | 0 | 0 | 0 | 0 |
| Farming (1) | 1 | 0 | 0 | 0 | 0 |
| Cryptocurrencies (1) | 0 | 0 | 0 | 1 | 0 |
| Internet (1) | 0 | 0 | 0 | 1 | 0 |
|  |  |  |  |  |  |
| **Grand Total (70)** | **37** | **21** | **3** | **8** | **1** |

### 4. Discussion and Recommendation

In this segment, we are going to discourse the results and responses of the research questions. Later we discuss the limitations of our research and recommend some future directions as well.

### 4.1. RQ1: What are the application divisions in which blockchain adoptions are being utilized or developed in Bangladesh?

Based on our analysed literature, this study outlined 13 application sub-domains where blockchain technology is being used. From **4.1.1 to 4.1.13**, this study will describe those sub-domains extensively and also justifies them with respect to Bangladesh.

**4.1.1. Governance**

Almost one-fourth (23) of the total articles considered (70) in this study are governance-related (Table-11). These articles include supply chain, e-voting, land registration, management system.

This article is about Driving License Supervision and Driver's Repute System for our country [23]. In that article the author tried to make a highlight on blockchain and smart contracts that created an improved driving license administration, driver's reputation framework, and computerized a few repetitive steps for a proficient system.

Another article is about E-voting by using blockchain technology [33]. In that kind of article an IoT based framework is planned to exchange information from e-voting gadgets to the nodes and the approach uses a conveyed ledger technology where information is shared and conveyed into a network.

Bangladesh is a country where corruption and mismanagement are major issues against development. Every year corruption and bad governance cost lots of money in Bangladesh[3] . If more research and a real implementation of blockchain technology can be done in Bangladesh, the country will be able to save millions.

**4.1.2. Healthcare**

From (70) articles there are only (6) articles regarding healthcare (Table-11). These articles are about secure data sharing about healthcare, Prediction of Dengue Infected Areas, Electronic Medical Records (EMR).

This article is regarding the Prediction of Dengue Infected Areas using the technology of the Crowdsourcing Framework [57]. By using this technology, we can search for the crowdsourcing and Dengue that can be easily spread to some kind of places and take the patients far from that place and use Ethereum-blockchain to recover the sanctuary of the background and avoid fake area information.

Another group of articles regarding blockchain-Based Information EMR in a Health care Communication Scheme here presents a blockchain-based show for securing EMR [76]. Utilized SHA256 protected hash calculation for creating a special and alike 256-bit or 32-byte hash rate for a specific therapeutic result.

The situation of the health sector in Bangladesh is not much good. There are corruption and mismanagement which reduce the hope of the patient to get proper treatments [93]. If it is probable to assure proper usage of blockchain technology it will beneficial for all. Possible implementation domains could be Covid-19 vaccine supply, medicine supply, specialist doctor information, data security of medical organizations.

**4.1.3. Security**

After Governance, the articles are mostly found which are related to security (Table-11). From (70) articles we find out (18) security-related articles. These articles are about Security management, Attack Detection, Examining Usability Issues, Data security, Secure Data Transaction.

First article is regarding the improvement of data security through blockchain, Like data access from cloud data storing by unapproved clients [34]. This exertion employs thumbprint as a biometric method and progressed encoding usual as a reliable encoding framework.

Another group of article is blockchain-Based Cryptocurrency Wallets [77]. This article investigates the common convenience topics with computer and cell or phone-based cases and the consequences uncover that mutual wallet need great convenience in executing the basic assignments which are going to be better quality. It essentially outlines the discoveries and point out the viewpoints where the issues exist.

---

[3] https://www.u4.no/publications/overview-of-corruption-and-anti-corruption-in-bangladesh-2019

In Bangladesh, blockchain-security related articles are quite familiar and that kind of technology is being used in different kinds of technological sectors also. For that reason, different kinds of crime are being prevented. Possible implementation domains could be secure mobile banking, data transfer.

**4.1.4. Privacy**

From (70) articles there are only (7) articles regarding Privacy in (Table-11). These articles are about Implementation for Smart Cities, Smart National Identity Card, Privacy-friendly platform for healthcare.

This article is related to blockchain-built SDN-IoT constructional security with NFV Execution for Canny Cities where planning for smart cities has been planned [27]. The architecture safeguards metadata within individual coating in addition to load the process.

Another article is about the Protected Smart National Identity Card Management proposal using blockchain [43]. Here in this article, most recent improvements in the field of NID card framework by implementing blockchain as a model where moreover talk about the working process of blockchain besides the contests confronted and upcoming points of view are also elaborated.

Bangladesh is a country where privacy can be unlocked very easily and the security system is not up to the mark[4]. The data is being stolen very easily by any person or any organization. We need to work on it.

**4.1.5. Survey**

We find out only (4) articles regarding surveys using blockchain technology (Table-11). These articles are about the challenges and needs of blockchain software, File System and Securely Manageable (EMR), Feedback on Blockchain Technology.

An article is related to a feedback system using blockchain technology [39]. In that article digital feedback system will be utilized where input will be guaranteed with the namelessness of the users and also make sure the belief of users is being stalled when feedback is given. Moreover, it'll grant a thought of how the people of an organization think about certain themes.

Another article is about software expansion applies of blockchain schemes [82], which is also an inspection-based article which investigates the program-building practices of these ventures. Computer program building strategies counting challenging and sanctuary best hones got to be adjusted with reality to discourse the exceptional features of blockchain.

Bangladesh is a developing country and needs to make more surveys about blockchain technology. "The survey articles are not very much available in our country", stated according to an official statement by the **Bangladesh National Bank**. From these survey-type articles, we can understand which technology is easy to be executed by exact end-user and which is beneficial for all. We need to work more on this.

**4.1.6. Review**

Regarding the review of sub-domains, there are only (2) articles to found in (**Table-11**). Those articles are about Healthcare applications and SDN and blockchain-based IoT Countermeasures.

The first review article is about developing blockchain technology in Health care Applications [75]. This considers methodically looks all-important explore articles about blockchain in the healthcare system in three available catalogues. They think about utilizing the characterized catchphrases 'blockchain', 'electronic health archives', 'health care and their varieties'.

Another article talks about approximately SDN and blockchain built IoT Countermeasures [50]. Here, watching the IoT arranged highlights, security details, and contests conjointly profound inquire about approximately SDN and blockchain-grounded which are tended for IoT to organize and also sanctuary with the scenario.

There is a lack of research into the use of blockchain technology. Researchers should focus on sub-domains such as medical and healthcare applications, Agricultural applications, Banking sector, etc. Students need to come forward to find new inventions regarding this technology.

**4.1.7. Data**

There are only (3) articles found in (Table-11), regarding data. Those articles are about Data Distribution, Data Sharing, and Remote Database Access.

---

[4] https://www.thedailystar.net/law-our-rights/urge-enact-privacy-and-data-protection-law-1527220

So, the first article is about blockchain Networks Using Lapping Information Delivery [90], and it will be reducing storage requirements. In this article, a strategy to separate the entire blockchain of contacts and a variety of duplicates has been proposed. At that point, these shards are distributed consistently over the nodes within the organization.

Another article is about the Addition of blockchain and Further Records to Get to the Protocol-Grounded Record [80]. This article recommends utilizing blockchain for fortifying small tries against hacking by disturbing the organization at any point of completing any changes to all the data without utilizing the official channels. It will be completed by the blockchain technology's innate hash repetition and withdrawal calculation.

The next article is regarding data security but data analysis articles are not sufficient in Bangladesh. Data is not much secure in our country. So, it is needed to research data analysis and statistics to make the data more secure and more useful. So possible implementation domains could be data analysis, data transfer security, and so on.

### 4.1.8. Storage

Only (1) article is being found by us regarding storage (Table-11). The only article is about Increasing Throughput and Reducing Storage.

So, this article is about Growing Quantity and Tumbling Storage Bloating Problem by IPFS and Dual-blockchain Method [31]. This Paper is persuaded due to the need for versatility on the prevailing blockchain innovations. A conveyed capacity framework IPFS is utilized to avoid putting absent obligations and to expand output.

There are very few articles and papers that are published in the storage sub-domain. In Bangladesh, storage-related papers and articles are not popular because the capability to maintain the storage and usability of data storage is not sufficient. Most of the web domains are being outsourced because storage is not sufficient. So, storage-related research and articles are not justified for our country. If the government take proper steps, storage-based domain-related work can be made a huge opportunity for blockchain technology.

### 4.1.9. Architecture

Like review, there are only (2) articles regarding architecture that we found in (Table-11). Those articles are about ASIC-resistant for blockchain systems and another one is Efficient blockchain Segmented Work.

So, the first article is about ASIC-resistant Proof-of-Work for blockchain systems [45]. This article proposed a Filtered Proof-of-Work (FPoW) and its' ASIC-resistivity is assessed to create an upcoming-resistant ASIC-resistant agreement convention.

Another article is about Efficient blockchain systems based on Proof of Segmented Work [32]. This system is attempted to decrease energy utilization by restricting the interest of all the hubs within the network by putting a parcel of emphasis on the concept of fair reward distribution which isn't considered in most cryptocurrencies.

Blockchain Architectural articles can improve the architectural situation of our country's data security. Every year lots of websites and data are being hacked and stolen by hackers. If the research regarding architecture is performed well the security will be increased enormously. For that, the Bangladeshi author should focus on blockchain architecture significantly. Thus, some possible implementation domains could be Hierarchical Blockchain, Edge Computing, and so on.

### 4.1.10. Information Verification

Like storage, there are only (1) article regarding Information Verification that we found in (Table-11). This article is about Fake News Detection.

The article is about Fake News Detection in communal media exploitation of blockchain [54]. This article is studied based on Fake news, which is more appealing than the genuine one. In this way, individuals got to be confused. Utilizing the preferences of blockchains peer-to-peer arranged concepts identify fake news in social media.

In our country Rumours is a common incidence. People take action without judging the right information. For that reason, there is a lot of harm and damage that has been done in our country. It's a political, or even become an economic issue sometimes. It will be justified, if Information verification-related blockchain papers and research-related works are focused properly in Bangladesh. So, possible implementation domains could be Facebook post verification, stock exchange news verification, etc.

**4.1.11. Farming**

Like storage and Information Verification, there are only (1) article regarding farming that had been found in (Table-11). This article is about the Unconventional Method of Farming without Mediators.

The only farming sub-domain related article is "An Unconventional Means of Farming Without Mediators through blockchain" [55]. This article is projected on a dispersed agricultural stage, named KHET to resolution the specified problems and establishes belief and regionalization between agronomic partners such as agronomists, supply corporations, and marketplaces.

Bangladesh is an agricultural country. 80 percent of the people of our country are directly connected and about 90-95 percent of people are indirectly connected with agriculture and farming. So, in this sector blockchain technology should grow up the research for our farmers and the interpreters. By using the technology, the production and the supply system of farming and agricultural product will be easy and the waste of products will get to be decreased. The possible implementation domain could be Farm Product Supply in this case.

**4.1.12. Cryptocurrencies**

Like storage and Information Verification and farming, there is only (1) article regarding Cryptocurrencies (Table-11).

This article is about Trustworthy Electronic voting by Familiar blockchain [64]. Electronic voting is a productive strategy that seems to offer assistance to individuals who seem not to vote for a restricted period. It too solves the issue of long voting time due to numerous voters or more complicated voting steps. So, proposing cryptanalysis on a dependable electronic voting plot with blockchain and discovers out that there are some issues in the main proposed plot.

A cryptocurrency is a digital medium of exchange such as the US Dollar. Cryptocurrency is not a regular thing in Bangladesh. Very little amount of paperwork has been done about "Cryptocurrency" in Bangladesh. So, more research is needed to execute this technology for Bangladesh.

**4.1.13. Internet**

Like storage and Information Verification and farming, there are also only (1) article found regarding Internet (Table-11). This article is about "Secured blockchain-Based Decentralised Internet".

This article proposed a new decentralized internet-based article. This article attempts to depict with blockchain technology into the decentralization of the web [66]. A decentralized arrangement assures to empower the web to function from the smart-phone or tablet of anyone rather than centralized servers.

In Bangladesh Internet is quite available now. Most of the person knows and uses it for a different purpose. But the internet service is not secure in our country. Most of the people got harassed on the platform of the internet. Females are mostly the victim of harassment. Blockchain technology can prevent such problems. So, we need to work and study internet based blockchain technology. One possible implementation domain could be Safe Internet Blockchain Security.

**4.2. RQ2: What are the impediments and encounters of blockchain technology adoption in Bangladesh?**

After a thorough analysis of several blockchain technology-related research articles as well as existing policies prevailing in Bangladesh, this study identifies major challenges and obstacles of this technology adoption in Bangladesh.

**4.2.1. Technical Challenges**

As an underdeveloped country where a limited amount of money is being dedicated to the research and development sector, Bangladesh faces several technical limitations. Although blockchain technology experience some common technical issues[5], how those limitations are affected here are mentioned herewith.

**Immaturity:** Blockchain is still immature because it is evolving and this is one of the major limitations of blockchain. For immaturity, blockchain cannot be used in a wide range of adoption. However, there are few domains in blockchain that are matured to be used in the production environment.

---

[5] https://www.finextra.com/blogposting/18496/remaining-challenges-of-blockchain-adoption-and-possible-solutions

**Lack of Standardization programming language:** To adopt blockchain, lack of standardization programming language is another problem. According to University Grants Commission (UGC), undergraduates' advice for making a typical Syllabus that is limited to C, C++, Java, C#, PHP, and python might not be enough to develop blockchain[6].

**Scalability:** As previously said, the scalability of major public blockchain schemes is inadequate. The private blockchain networks seem to provide greater scalability in principle. However, since their scalability is not well known, there is no definitive evidence of it. For any large-scale adoption, this is a long way away.

**Energy Scarcity:** In developing countries like Bangladesh, the energy crisis is a major problem. Consensus algorithms (such as Proof of Work) that are used in big public blockchain networks use a lot of power. According to the Bitcoin network's current energy usage is about 113.47 TWH, which is equal to Netherland's energy consumption [94]. Because of the simplicity of the consensus algorithm used in Bitcoin, the rate of energy use rises as Bitcoin's popularity expands. As a result, many people are unsure if this is likely to continue in the future. So, energy scarcity is a big problem to adopt blockchain in Bangladesh.

**Internet Scarcity:** Internet Scarcity is another major problem in Bangladesh. Rendering the Bangladesh Telecommunication Regulatory Commission's (BTRC) figures, the country's net customers totalled 108.188 million in August 2020, up from 54.120 million after December 2015 [95]. Customers have increased rapidly but our capacity is not. Bangladesh is in the 98th position out of 175 nations (speed).

**Security and Vulnerability in Internet Infrastructure:** According to Bangladesh Telecommunication Regulatory Commission, active internet users were 116.14 million in March 2021[7]. Many national and global corporations are facilitating online shopping, banking, communication, and a variety of other e-commerce services. But the matter of concern is that most of the software used in our country is pirated. Taking advantage of this, criminals infiltrate digital networks and engage in illegal activities like phishing, hacking, and stealing personal data and institutional information.

**High Implementation Cost**: Implementation of blockchain costs a lot. Because blockchain is a feature-dependent technology, the ultimate pricing will vary contingent on the scheme stipulations. We must mention that the charge of emerging a blockchain application starts from $5,000 to $200,000 that is too much for Bangladesh. It is also one of the obstacles of developing blockchain technology.

**Compatibility with already established systems:** There are many obstacles to adopt blockchain technology in Bangladesh and one of the causes is compatibility with the existing system. To apply blockchain we are in lack of expertise, advanced technology, and so on. Moreover, all the people in Bangladesh are still not aware of blockchain technology. They don't know what is it, how to use it. It feels like "an already established system is much easier than blockchain." For that, it is necessary to adjust to an already established system.

**Less opportunity for using paid software:** As a developing country, Bangladesh is not technologically advanced enough. The implementation cost of blockchain is much higher in our country. The rate of piracy in our county is 92%[8]. Using paid software cost a lot because generally, we have to buy the software from abroad since we are yet not much developed in software industries. In addition, weak law enforcement, and the lacking income, moral degradation, and the lack of awareness of lawbreaking is also the cause.

### 4.2.2. Organizational Challenge

**Lack of awareness:** Many companies are unaware of the benefits that private blockchain technologies will provide to their operations. Many people believe in cryptocurrency to be the only usage case for blockchain, so they treat the two words interchangeably. We can see from the result and analysis only (75) articles on (13) sectors in Bangladesh used blockchain technology.

**Lack of regulatory mechanism:** Adopters gain hope from a regulatory framework covering any novel technology. The lack of such a structure, on the other hand, confuses. From our analysis, Bangladesh has no such regulatory strategy. In this paper [96], we find that the law and regulatory policy because it could impact how far and how fast the technology could develop. This also balances the system risk.

**Legal Concerns about Policy:** The Administration of the People's Republic of Bangladesh released the National blockchain Strategy: "Pathway to become a blockchain-enabled Nation in 2020." But like a few other countries, Bangladesh also banned any sort of cryptocurrencies in 2017. It can be said hostile view towards cryptocurrencies. Another sector except financial blockchain growth is slow. So, Bangladesh government need to change their policy towards blockchain and make a policy that benefits everyone.

---

[6] http://www.ugc.gov.bd/site/view/policies/-
[7] http://www.btrc.gov.bd/content/internet-subscribers-bangladesh-march-2021
[8] https://www.nationmaster.com/country-info/stats/Crime/Software-piracy-rate

**Governance:** Every blockchain system's governance is a key problem that is commonly ignored. In several cases, the governance of a blockchain determines its success and adoption. Surprisingly, various facets of blockchain governance are analogous to how a company, consortium, or group of companies distributes responsibility within themselves. Before deciding to use blockchain for their solutions, each company must consider this. This necessitates a modern governance paradigm, which may be difficult for any organization to implement. Many organizations will benefit greatly from a controlled blockchain network (managed by the government or any private organization) since they would not have to think about governance issues [97].

#### 4.2.3. Human Resource Challenges

As a new technology, there is a lack of human resources such as a team of blockchain experts and technologists. Our universities and college academicians are very few who are researching on blockchain and published their work on blockchain. To overcome this, we need government support.

**Lack of technological expertise:** Any new technology faces a skills gap, particularly during its early stages of development. Blockchain technology is no exception; blockchain professionals are in short supply all over the world. This is especially true in Bangladesh as a developing country in South Asia. Furthermore, there is a lack of understanding of Blockchain and its benefits among the general public.

### 4.3. RQ3: How Policymakers should approach to adopt the blockchain technology for sustainable development in Bangladesh?

Finally, this study could find some strategies that must be taken by the government, researchers, and educationists for sustainable development of Bangladesh.

#### 4.3.1. Government and Policy-Makers

**Firstly,** create research funds for researchers and educational institutes to build infrastructure.
**Secondly,** organize national and international seminars and workshops.
**Thirdly,** remove policy-oriented barriers in the path to expand blockchain, cryptocurrencies, and Non-Fungible Token (NFT).
**Fourthly,** involve more researchers, teachers, and stakeholders to form strategies with politicians.

#### 4.3.2. Researchers

**Firstly**, researchers must collaborate with international peers, government, and educational institutes to form local research institutes that could be solely focused on blockchain technology.
**Secondly**, researchers must focus on local priorities (i.e., farming, corruption, education, etc.) precisely while developing blockchain-assisted systems.

#### 4.3.3. Educational Institutes

**Firstly,** all the educational institutes must arrange more and more seminars on blockchain technology. In addition, basic blockchain technology and cryptocurrency can also be incorporated in undergraduate and graduate-level studies [98], which perhaps are hardly practicing in any of Bangladeshi Universities in current time [99].

**Secondly,** the educational institutions may arrange blockchain and cryptocurrency-related in-house competitions, so that the students are encouraged about executing blockchain in their day-to-day life.

## 5. Conclusion:

This study successfully outlined those Blockchain technology-related articles which are published by the Bangladeshi authors in the five most prestigious databases such as IEEE Xplore, ACM, ScienceDirect, Taylor & Francis, and SpringerLink. By doing the manual evaluation of these articles from those five databases, we separated them into three major domains and thirteen sub-domains. By doing the manual evaluation, we unleashed 70 articles which are Blockchain technology-related articles and also published by the Bangladeshi authors. Moreover, we found out the key concepts for each of those articles.

Furthermore, by doing that, we found out that most published articles that are based on the application on governance, security, and privacy-related articles. It also helped us to find out those sectors that are currently being worked on Blockchain Technology and also needs more future scopes to use the technology for the betterment irrespective of the health, data, and architecture of the Bangladeshi people. These kinds of articles are not sufficient in numbers. So Future Bangladeshi researchers have lots of opportunities to work on these sectors.  Besides that, this study also outlined the future challenges, directions, and recommendation-strategies for local government, educational institutions, and researchers. In the end, this study concludes that more investment and hands-on experience from the private and public sectors will surely be increased through the real-world deployment of blockchain technology in Bangladesh.

**Acknowledgement**

The authors would like to thank American International University-Bangladesh (AIUB) authority for providing technical supports. In addition, we would like to pay our sincere gratitude to the anonymous reviewer and editor for providing valuable suggestions during the review period.